# Electric-field-modulated topological phase transition in AlSb/InSe heterobilayers

D. Q. Fang[1*], D. W. Wang[2,3]

[1]MOE Key Laboratory for Nonequilibrium Synthesis and Modulation of Condensed Matter, School of Physics, Xi'an Jiaotong University, Xi'an 710049, China

[2]School of Microelectronics & State Key Laboratory for Mechanical Behavior of Materials, Xi'an Jiaotong University, Xi'an 710049, China

[3]Key Lab of Micro-Nano Electronics and System Integration of Xi'an City, Xi'an Jiaotong University, Xi'an 710049, China

Abstract

Searching for controllable topological phase by means of external stimuli in two-dimensional (2D) material-based van der Waals (vdW) heterostructures is currently an active field for both the underlying physics and practical applications. Here, using first-principles calculations, we investigate electric-field-modulated topological phase transition in a vdW heterobilayer formed by vertically stacking 2D AlSb and InSe monolayers. The AlSb/InSe heterobilayer studied possesses both dynamical and thermal stabilities, which is a direct bandgap semiconductor and forms a Z-scheme heterojunction. With inclusion of spin-orbit coupling (SOC) and applying external electric field, the bandgap decreases at first and then increase, and a trivial insulator to topological insulator phase transition is observed. For the topological insulator phase, band inversion is ascribed to the strong SOC of p orbitals of Sb. Our work paves the way for the design and application of multifunctional nanoscale devices such as topological field effect transistor.

*fangdqphy@xjtu.edu.cn

# I. Introduction

Since the discovery of graphene [1], many other two-dimensional (2D) materials have been realized in experiments by mechanical exfoliation [2] or molecular beam epitaxy. They exhibit novel electronic, magnetic, and optical properties. For example, monolayer 1T'-WTe$_2$ was proposed as a quantum spin Hall (QSH) insulator with large bandgap [3], as evidenced by the angle-resolved photoemission and scanning tunnelling spectroscopy measurements [4]; monolayer CrI$_3$ was found to be an Ising ferromagnet with out-of-plane spin orientation and a Curie temperature of 45 kelvin [5]; a valley polarization of 30% in pristine monolayer MoS$_2$ was achieved by optical pumping with circularly polarized light [6].

In parallel with the efforts on monolayer materials, van der Waals (vdW) heterostructures made by vertically stacking different 2D crystals have attracted considerable interest [7], offering remarkable platform to study new physical phenomena and a huge variety of candidates for applications. Cao et al. observed tunable zero-resistance states with a critical temperature of up to 1.7 kelvin in bilayer graphene with twist angles of about 1.1° and an electrostatic doping of the material away from the correlated insulating states, making twisted bilayer graphene an ideal material for investigations of strongly correlated phenomena [8]. Ubrig et al. observed robust k-direct interlayer transitions at Γ in the InSe/WS$_2$ vdW interfaces with momentum-matched type-II band alignments, irrespective of a substantial lattice constant mismatch between the constituents and whether the constituent materials are direct or indirect bandgap semiconductors [9].

Several vdW heterostructures were investigated using first-principles calculations to realize controllable QSH states. Previous works demonstrated that the free-standing Bi(111)-bilayer is a 2D topological insulator [10]. Recently, Bai et al. showed that in the Bi(111)-bilayer/α-In$_2$Se$_3$ heterostructures the transition from a topological insulator to a trivial metal can be realized by reversing the electric polarization of the ferroelectric α-In$_2$Se$_3$ [11]. Another strategy was proposed based on vdW heterobilayers comprising a ferroelectric In$_2$Se$_3$ layer and a trivial 2D insulator such as monolayer Sb and monolayer CuI [12,13]. In one polarization state the system is a trivial semiconductor, while it becomes a topological insulator with a sizable bandgap upon polarization reversal, arising from a combination of intralayer spin-orbit coupling and interlayer hybridization [13].

Monolayer InSe possesses a honeycomb atomic arrangement produced by four planes of atoms, Se-In-In-Se, which has been fabricated by mechanically exfoliation and shows high environmental stability and superior electronic and optical properties [14,15]. Monolayer AlSb, predicted to be stable in a double layer honeycomb structure [16], was recently synthesized by molecular beam epitaxy growth [17] and proposed as a candidate for the realization of excitonic insulator [18]. In view of the rich physical properties of vdW heterostructures, in this work, we construct a heterobilayer composed of AlSb and InSe monolayers and systematically investigate the stacking configuration, stability, electronic structure, and effect of external electric field. We find that AlSb/InSe heterobilayer is a trivial semiconductor with the bottom of the conduction band and the top of the valence band located at the Γ point and forms a Z-scheme heterojunction, while it becomes a topological insulator under external electric field.

## II. Computational details

All first-principles calculations are performed within the framework of density functional theory (DFT), using the projector augmented-wave method [19], as implemented in the Vienna Ab initio Simulation Package (VASP) [20]. The exchange-correlation potential is described by the generalized gradient approximation of Perdew, Burke, and Ernzerhof (PBE) [21]. The DFT-D3 method [22] is used to describe the van der Waals interactions. We use a Γ-centered k-point grid of 11×11×1 to sample the Brillouin zone, and a 500-eV cutoff energy for the plane wave basis set. In the out-of-plane direction a vacuum region of more than 12 Å is employed to avoid the interactions between the neighboring images and the dipole correction is exerted [23]. Convergence criterion for the electronic self-consistency is set to $10^{-6}$ eV and $10^{-8}$ eV for structural relaxation and static calculation, respectively. Lattice constants and atomic positions are both optimized until the forces are less than 0.005 eV/Å. Since PBE functional tends to underestimate the bandgap, we employ Heyd−Scuseria−Ernzerhof (HSE06) [24] hybrid functional to calculate the band structure. We find that taking into account 35% Hartree-Fock (HF) exchange produces accurate positions of the band edges for InSe and AlSb monolayers compared to the quasiparticle $G_0W_0$ calculations (see Figure S1), and therefore HSE06 functional with 35% HF exchange is used for the electronic structure calculations of AlSb/InSe heterobilayers. The tight-binding model is obtained by projecting onto In-s,$p_z$,$p_x$,$p_y$, Se-$p_z$,$p_x$,$p_y$, Al-s,$p_z$,$p_x$,$p_y$, and Sb-$p_z$,$p_x$,$p_y$ orbitals using Wannierization procedure as implemented in the Wannier90 code [25]. Hybrid functional band structures with spin-orbit coupling (SOC) are calculated via Wannier interpolation [26].

Calculations for phonon dispersion are carried out using the finite difference method implemented in the Phonopy code [27] and 200-atom 5×5 supercells. Ab initio molecular dynamics (MD) simulation at 300 K in the canonical ensemble is performed using a 4×4 supercell, a Γ k point, and a 400-eV cutoff energy.

### III. Results and discussion

A. Stacking configuration and structural stability

We first investigate the properties of InSe and AlSb monolayers. The optimized lattice constant is predicted as *a* = 4.05 Å for monolayer InSe and *a* = 4.29 Å for monolayer AlSb, in good agreement with preceding studies [28,29]. Hybrid functional (with 35% HF exchange) calculations show that monolayer InSe is an indirect bandgap semiconductor with a bandgap of 2.66 eV with the conduction band minimum (CBM) at the Γ point and the valence band maximum (VBM) along the Γ to K direction, and monolayer AlSb is a direct bandgap semiconductor with a bandgap of 0.94 eV with both the CBM and VBM at the Γ point (Figure S2). With inclusion of SOC, the bandgap of monolayer InSe changes slightly, being 2.61 eV, while that of monolayer AlSb decreases substantially to 0.73 eV. For monolayer AlSb, SOC lifts the degeneracy of the two highest valence bands at the Γ point dominated by the $p_x$ and $p_y$ orbitals of Sb and the split bands lead to a large gap reduction [Figure S2(d)].

Next, we consider AlSb/InSe heterobilayers by vertically stacking AlSb and InSe monolayers. Figure 1 shows the four possible stacking configurations AA, AB, AB', and AA'. The relaxed lattice constants of the four configurations are same, i.e. 4.20 Å, while the relaxed interlayer distances for the AA, AB, AB', and AA' configurations are 3.21 Å, 3.07 Å, 3.07 Å, and 3.17 Å, respectively. To examine the stabilities of these configurations, the binding energy of the interface is calculated as $E_b = E_{AlSb} + E_{InSe} - E_{AlSb/InSe}$, where $E_{AlSb}$, $E_{InSe}$, and $E_{AlSb/InSe}$ are the total energies per cell of monolayer AlSb, monolayer InSe, and AlSb/InSe heterobilayer, respectively. The obtained binding energies are 173.4, 181.8, 184.1, and 175.3 meV/cell (equivalent to 11.35, 11.90, 12.05, 11.47 meV/Å$^2$) for the AA, AB, AB', and AA' configurations, respectively, which are in accord with previously reported values for weakly bonded layered compounds (in the energy interval of 13-21 meV/Å$^2$) [30]. The relatively small interlayer binding energies indicate the domination of the vdW interaction for AlSb/InSe heterobilayers. Our calculations also confirm that AB' is the ground state configuration with AA, AB, and AA' higher in total energy per cell by 10.7 meV, 2.3 meV, and 8.8 meV, respectively.

Figure 2(a) shows the calculated phonon dispersion of AlSb/InSe heterobilayer in the ground state AB' configuration. There is no imaginary frequency along the path connecting high symmetry points in the Brillouin zone, except for a small pocket around the Γ point resulting from the Fourier interpolation. This small pocket is not a real physical effect but reflects the difficulty of achieving numerical convergence for the flexural phonon branch of 2D materials. Thus, the phonon dispersion indicates that AB'-stacked AlSb/InSe heterobilayer is dynamically stable. To check the thermal stability, we perform MD simulation at 300K. There is small energy fluctuation during the simulation [Figure 2(b)] and the framework of heterobilayer after 10 ps is well-kept [Figure S3(b)], demonstrating the thermal stability of this system.

To examine the change of charge at the interface, we calculate the charge density difference between the AB'-stacked AlSb/InSe heterobilayer and isolated monolayers, i.e. $\Delta\rho = \rho_{AlSb/InSe} - \rho_{AlSb} - \rho_{InSe}$, as shown in Fig. 2(c). It is visible that interfacial interaction leads to the depletion (accumulation) of electrons in the region near monolayer AlSb (InSe), resulting in the formation of a built-in electric field at the interface along the direction from AlSb layer to InSe layer.

B. Electronic structure

Figure 3(a) shows the calculated electronic band structure of AB'-stacked AlSb/InSe heterobilayer using hybrid functional without SOC. The AB'-stacked AlSb/InSe heterobilayer exhibits a direct bandgap of 0.40 eV at the Γ point. The VBM is degenerate and mainly contributed by the $p_x$ and $p_y$ orbitals of Sb, while the CBM is dominated by InSe layer, especially its s and $p_z$ orbitals from In and Se, respectively, as shown in Figure S4. The AA-, AB-, and AA'-stacked AlSb/InSe heterobilayers have similar band structure, with direct bandgaps of 0.36, 0.41, and 0.38 eV at the Γ point, respectively, as shown in Figure S5. Since the built-in electric field at the interface points from AlSb to InSe, the AB'-stacked AlSb/InSe heterobilayer forms a Z-scheme heterojunction [31], which facilitates interlayer electron-hole recombination and thus hinders the migration of photogenerated holes (electrons) between the VBMs (CBMs) of different materials, maintaining a high redox ability. For the AB'-stacked AlSb/InSe heterobilayer, the CBM of AlSb layer is estimated to be -3.5 eV and the VBM of InSe layer -6.1 eV relative to vacuum, which straddle water redox potentials, making its use in optoelectronics and photocatalysis possible.

Once SOC is accounted for, the degenerated $|p_x, p_y>$ states of Sb at the $\Gamma$ point around the Fermi level are split into the $|p_{x+iy,\uparrow}, p_{x-iy,\downarrow}>$ and $|p_{x-iy,\uparrow}, p_{x+iy,\downarrow}>$ states with a splitting energy of 0.42 eV, similar to the isolated AlSb monolayer, reducing the bandgap to 0.20 eV, as shown in Figs. 3(b) and 3(f).

C. Effect of external electric field

External electric field is one of the effective means to tune the materials' properties. For example, mono- and bilayer $Na_3Bi$ are QSH insulators, whereas an electric field tunes their phases from topological insulator to conventional insulator due to a Stark-effect-driven transition [32]. Herein, we investigate how the band structure of AlSb/InSe heterobilayer changes with applying electric field. Figure 3(c) presents the band structure of AB'-stacked AlSb/InSe heterobilayer under an external electric field $E_{ext}$ = 0.4 V/Å without SOC and the direction of external electric field is shown in Fig. 3(e). We find that the bandgap of AlSb/InSe heterobilayer decreases significantly to 0.072 eV. The VBM is still contributed by the $p_x$ and $p_y$ orbitals of Sb and the CBM by s and $p_z$ orbitals from In and Se, respectively.

Upon inclusion of SOC a band inversion between VBM and CBM appears, as shown in Fig. 3(d), and the gap at the $\Gamma$ point around the Fermi level is 0.062 eV and the whole bandgap of the system is 0.033 eV. We track the origin of the band inversion: the SOC strength of p orbitals of Sb $\lambda_{Sb}$ is varied from 0 to 0.4 eV based on a tight-binding model (APPENDIX). For $\lambda_{Sb}$ = 0.4 eV, a band inversion is observed and an energy gap is opened, similar to DFT calculations. The SOC of other elements has negligible effect on the band inversion. This proves that the SOC of p orbitals of Sb indeed causes the band inversion, a feature of nontrivial electronic state. Figure 3(g) shows the orbitals' evolution at the Γ point under $E_{ext}$ = 0.4 V/Å. The $|p_{x+iy,\uparrow}, p_{x-iy,\downarrow}\rangle$ state of Sb lies above the Fermi level due to SOC and the states from InSe layer below the Fermi level, showing an inverted gap.

To determine the topological phase, we calculate the hybrid Wannier charge center (WCC) [33] defined as

$$WCC_n(\vec{k_2}) = \frac{ia_1}{2\pi} \int d\vec{k_1} \left\langle u_{n,\vec{k_1},\vec{k_2}} | \partial_{\vec{k_1}} | u_{n,\vec{k_1},\vec{k_2}} \right\rangle,$$

where $u_{n,\vec{k_1},\vec{k_2}}$ is the lattice-periodic part of the Bloch function, using the WannierTools code [34]. The $\mathbb{Z}_2$ invariant can be deduced from the WCC evolution. When the WCC curves are crossed by an arbitrary horizontal line an odd number of times, the whole system is topologically nontrivial, $\mathbb{Z}_2$ = 1 [35]. Figures 4(a) and 4(c) show the calculated evolution of WCC for the AB'-stacked AlSb/InSe heterobilayer in the absence of external electric field and under $E_{ext}$ = 0.4 V/Å, respectively. The number of WCC crossings is an even integer for the case without external electric field, yielding a trivial topological invariant $\mathbb{Z}_2$ = 0. However, for the case under $E_{ext}$ = 0.4 V/Å the odd times WCC crossings are found, indicating a topologically nontrivial phase, $\mathbb{Z}_2$ = 1, and the presence of the QSH effect. As a result of nontrivial topology, we expect the presence of helical edge states that cross the bulk gap. Figures 4(b) and 4(d) show the calculated edge spectra using the iterative Green's function approach as implemented in the WannierTools code [34]. There are gapless edge states inside the bulk gap that connect the conduction bands with valence bands for the AB'-stacked AlSb/InSe heterobilayer under $E_{\text{ext}}$ = 0.4 V/Å [Fig. 4(d)], demonstrating that it is a topological insulator.

Finally, we investigate the change of the bulk bandgap of AB'-stacked AlSb/InSe heterobilayer with increasing the strength of external electric field. The heterobilayer studied has a large thickness (12.3 Å) and the electronic states are quantum confined in different layers. When electric field is applied along the confinement direction, an obvious change of the electronic states can take place, leading to a dramatic variation of the bandgap, referred to as the quantum-confined Stark effect. As expected, the bandgap decreases almost linearly with increasing the strength of external electric field in the case of no SOC (Fig. 5), showing a giant Stark effect. When SOC is turned on, however, the bandgap decreases at first and then increases in the electric field range from 0 to 0.5 V/Å and the transition occurs at $E_{\text{ext}}$ = 0.31 V/Å. Our calculations show that AB'-stacked AlSb/InSe heterobilayer is in the topological insulator phase when $E_{\text{ext}}$ ≥ 0.31 V/Å whereas in the topologically trivial state when $E_{\text{ext}}$ < 0.31 V/Å. The topological gap can be improved from 0.033 eV under $E_{\text{ext}}$ = 0.4 V/Å to 0.041 eV under $E_{\text{ext}}$ = 0.5 V/Å, providing the potential for the application in the room-temperature topological field effect transistor.

## IV. Conclusions

In this work, we investigate the stability, electronic structure, and effect of external electric field for the AlSb/InSe heterobilayers using first-principles calculations. The AlSb/InSe heterobilayer with the AB' stacking configuration is energetically most favorable, possessing both dynamical and thermal stabilities, and is a direct bandgap semiconductor and forms a Z-scheme heterojunction due to the built-in electric field. With inclusion of SOC and applying external electric field, the bandgap decreases at first and then increase, and a trivial insulator to topological insulator phase transition is observed. For the topological insulator phase, band inversion is driven by the SOC of p orbitals of Sb. Our work provides a scheme for the realization of QSH insulator in topologically trivial vdW heterostructures and lays the foundations for the application of the AlSb/InSe heterobilayers in photocatalysis and topological field effect transistor.

**Supporting Material**

Comparison of the positions of band edges of InSe and AlSb monolayers relative to vacuum calculated using different methods; Electronic band structures of InSe and AlSb monolayers; MD simulation and VBM and CBM states for the AB'-stacked AlSb/InSe heterobilayer; Hybrid functional band structure without SOC for the AA, AB, AB', and AA' stacking configurations.

**ACKNOWLEDGMENTS**

We acknowledge the financial support from the National Natural Science Foundation

of China (Grant Nos. 11604254, 11974268, and 12111530061) and the Natural Science Foundation of Shaanxi Province (Grant No. 2019JQ-240). We also acknowledge the HPCC Platform of Xi'an Jiaotong University for providing the computing facilities.

**Conflict of interest**

The authors have no conflicts to disclose.

**APPENDIX**: Evolution of the band structure with tuning the SOC strength of p orbitals of Sb

The tight-binding Hamiltonian for the AB'-stacked AlSb/InSe heterobilayer under $E_{ext}$ = 0.4 V/Å without SOC $H_{TB}$ is constructed using the Wannier90 code based on the maximally-localized Wannier functions [25]. The SOC Hamiltonian is obtained by adding the on-site $H_{SOC} = \sum_i \lambda_i \mathbf{L}_i \cdot \mathbf{S}_i$ term [36], i.e., $H = H_{TB} + H_{SOC}$. Figure 6 shows the evolution of the band structure with tuning the SOC strength of p orbitals of Sb ($\lambda_{Sb}$). For $\lambda_{Sb}$= 0.4 eV, the bands around the Fermi level become inverted and the system is topologically nontrivial. The influence of the SOC of other elements on the band inversion is negligible.

References:

[1] K. S. Novoselov, A. K. Geim, S. V. Morozov, D. Jiang, Y. Zhang, S. V. Dubonos, I. V. Grigorieva, and A. A. Firsov, *Electric Field Effect in Atomically Thin Carbon Films*, Science **306**, 5696 (2004).

[2] Y. Huang, Y.-H. Pan, R. Yang, L.-H. Bao, L. Meng, H.-L. Luo, Y.-Q. Cai, G.-D. Liu, W.-J. Zhao, Z. Zhou, L.-M. Wu, Z.-L. Zhu, M. Huang, L.-W. Liu, L. Liu, P. Cheng, K.-H. Wu, S.-B. Tian, C.-Z. Gu, Y.-G. Shi, Y.-F. Guo, Z. G. Cheng, J.-P. Hu, L. Zhao, G.-H. Yang, E. Sutter, P. Sutter, Y.-L. Wang, W. Ji, X.-J. Zhou, and H.-J. Gao, *Universal Mechanical Exfoliation of Large-Area 2D Crystals*, Nat. Commun. **11**, 1 (2020).

[3] X. Qian, J. Liu, L. Fu, and J. Li, *Quantum Spin Hall Effect in Two-Dimensional Transition Metal Dichalcogenides*, Science **346**, 1344 (2014).

[4] S. Tang, C. Zhang, D. Wong, Z. Pedramrazi, H.-Z. Tsai, C. Jia, B. Moritz, M. Claassen, H. Ryu, S. Kahn, J. Jiang, H. Yan, M. Hashimoto, D. Lu, R. G. Moore, C.-C. Hwang, C. Hwang, Z. Hussain, Y. Chen, M. M. Ugeda, Z. Liu, X. Xie, T. P. Devereaux, M. F. Crommie, S.-K. Mo, and Z.-X. Shen, *Quantum Spin Hall State in Monolayer 1T'-WTe$_2$*, Nat. Phys. **13**, 7 (2017).

[5] B. Huang, G. Clark, E. Navarro-Moratalla, D. R. Klein, R. Cheng, K. L. Seyler, D. Zhong, E. Schmidgall, M. A. McGuire, D. H. Cobden, W. Yao, D. Xiao, P. Jarillo-Herrero, and X. Xu, *Layer-Dependent Ferromagnetism in a van Der Waals Crystal down to the Monolayer Limit*, Nature **546**, 7657 (2017).

[6] H. Zeng, J. Dai, W. Yao, D. Xiao, and X. Cui, *Valley Polarization in MoS$_2$ Monolayers by Optical Pumping*, Nat. Nanotechnol. **7**, 8 (2012).

[7] A. K. Geim and I. V. Grigorieva, *Van Der Waals Heterostructures*, Nature **499**, 7459 (2013).

[8] Y. Cao, V. Fatemi, S. Fang, K. Watanabe, T. Taniguchi, E. Kaxiras, and P. Jarillo-Herrero, *Unconventional Superconductivity in Magic-Angle Graphene Superlattices*, Nature **556**, 7699 (2018).

[9] N. Ubrig, E. Ponomarev, J. Zultak, D. Domaretskiy, V. Zólyomi, D. Terry, J. Howarth, I. Gutiérrez-Lezama, A. Zhukov, Z. R. Kudrynskyi, Z. D. Kovalyuk, A. Patané, T. Taniguchi, K. Watanabe, R. V. Gorbachev, V. I. Fal'ko, and A. F. Morpurgo, *Design of van Der Waals Interfaces for Broad-Spectrum Optoelectronics*, Nat. Mater. **19**, 3 (2020).

[10] Z. Liu, C.-X. Liu, Y.-S. Wu, W.-H. Duan, F. Liu, and J. Wu, *Stable Nontrivial Z$_2$ Topology in Ultrathin Bi (111) Films: A First-Principles Study*, Phys. Rev. Lett. **107**, 136805 (2011).

[11] H. Bai, X. Wang, W. Wu, P. He, Z. Xu, S. A. Yang, and Y. Lu, *Nonvolatile Ferroelectric Control of Topological States in Two-Dimensional Heterostructures*, Phys. Rev. B **102**, 235403 (2020).

[12] J.-J. Zhang, D. Zhu, and B. I. Yakobson, *Heterobilayer with Ferroelectric Switching of Topological State*, Nano Lett. **21**, 785 (2021).

[13] A. Marrazzo and M. Gibertini, *Twist-Resilient and Robust Ferroelectric Quantum Spin Hall Insulators Driven by van Der Waals Interactions*, Npj 2D Mater. Appl. **6**, 30 (2022).

[14] D. A. Bandurin, A. V. Tyurnina, G. L. Yu, A. Mishchenko, V. Zólyomi, S. V. Morozov, R. K. Kumar, R. V. Gorbachev, Z. R. Kudrynskyi, S. Pezzini, Z. D. Kovalyuk, U. Zeitler, K. S. Novoselov, A. Patanè, L. Eaves, I. V. Grigorieva, V. I. Fal'ko, A. K. Geim, and Y. Cao, *High Electron Mobility, Quantum Hall Effect and Anomalous Optical Response in Atomically Thin InSe*, Nat. Nanotechnol. **12**, 3 (2017).

[15] M. J. Hamer, J. Zultak, A. V. Tyurnina, V. Zólyomi, D. Terry, A. Barinov, A. Garner, J. Donoghue, A. P. Rooney, V. Kandyba, A. Giampietri, A. Graham, N. Teutsch, X. Xia, M. Koperski, S. J. Haigh, V. I. Fal'ko, R. V. Gorbachev, and N. R. Wilson, *Indirect to Direct Gap Crossover in Two-Dimensional InSe Revealed by Angle-Resolved Photoemission Spectroscopy*, ACS Nano **13**, 2136 (2019).

[16] M. C. Lucking, W. Xie, D.-H. Choe, D. West, T.-M. Lu, and S. B. Zhang, *Traditional Semiconductors in the Two-Dimensional Limit*, Phys. Rev. Lett. **120**, 8 (2018).

[17] L. Qin, Z.-H. Zhang, Z. Jiang, K. Fan, W.-H. Zhang, Q.-Y. Tang, H.-N. Xia, F. Meng, Q. Zhang, L. Gu, D. West, S. Zhang, and Y.-S. Fu, *Realization of AlSb in the Double-Layer Honeycomb Structure: A Robust Class of Two-Dimensional Material*, ACS Nano **15**, 5 (2021).

[18] S. Dong and Y. Li, *Excitonic Instability and Electronic Properties of AlSb in the Two-Dimensional Limit*, Phys. Rev. B **104**, 8 (2021).

[19] P. E. Blöchl, *Projector Augmented-Wave Method*, Phys. Rev. B **50**, 24 (1994).

[20] G. Kresse and J. Furthmüller, *Efficient Iterative Schemes for* Ab Initio *Total-Energy Calculations Using a Plane-Wave Basis Set*, Phys. Rev. B **54**, 16 (1996).

[21] J. P. Perdew, K. Burke, and M. Ernzerhof, *Generalized Gradient Approximation Made Simple*, Phys. Rev. Lett. **77**, 18 (1996).

[22] S. Grimme, J. Antony, S. Ehrlich, and H. Krieg, *A Consistent and Accurate* Ab Initio *Parametrization of Density Functional Dispersion Correction (DFT-D) for the 94 Elements H-Pu*, J. Chem. Phys. **132**, 15 (2010).

[23] J. Neugebauer and M. Scheffler, *Adsorbate-Substrate and Adsorbate-Adsorbate Interactions of Na and K Adlayers on Al(111)*, Phys. Rev. B **46**, 24 (1992).

[24] A. V. Krukau, O. A. Vydrov, A. F. Izmaylov, and G. E. Scuseria, *Influence of the Exchange Screening Parameter on the Performance of Screened Hybrid Functionals*, J. Chem. Phys. **125**, 22 (2006).

[25] A. A. Mostofi, J. R. Yates, G. Pizzi, Y.-S. Lee, I. Souza, D. Vanderbilt, and N. Marzari, *An Updated Version of Wannier90: A Tool for Obtaining Maximally-Localised Wannier Functions*, Comput. Phys. Commun. **185**, 8 (2014).

[26] N. Marzari, A. A. Mostofi, J. R. Yates, I. Souza, and D. Vanderbilt, *Maximally Localized Wannier Functions: Theory and Applications*, Rev. Mod. Phys. **84**, 1419 (2012).

[27] A. Togo and I. Tanaka, *First Principles Phonon Calculations in Materials Science*, Scr. Mater. **108**, 1 (2015).

[28] M. Wu, J. Shi, M. Zhang, Y. Ding, H. Wang, Y. Cen, and J. Lu, *Enhancement of Photoluminescence and Hole Mobility in 1- to 5-Layer InSe Due to the Top Valence-Band Inversion: Strain Effect*, Nanoscale **10**, 24 (2018).

[29] M. A. Mohebpour and M. B. Tagani, *First-Principles Study on the Electronic and Optical Properties of AlSb Monolayer*, Sci. Rep. **13**, 9925 (2023).

[30] T. Björkman, A. Gulans, A. V. Krasheninnikov, and R. M. Nieminen, *Van Der Waals Bonding in Layered Compounds from Advanced Density-Functional First-Principles Calculations*, Phys. Rev. Lett. **108**, 23 (2012).

[31] Q. Zhang, Y. Xiong, Y. Gao, J. Chen, W. Hu, and J. Yang, *First-Principles High-Throughput Inverse Design of Direct Momentum-Matching Band Alignment van Der Waals Heterostructures Utilizing Two-Dimensional Indirect Semiconductors*, Nano Lett. **24**, 3710 (2024).

[32] J. L. Collins, A. Tadich, W. Wu, L. C. Gomes, J. N. B. Rodrigues, C. Liu, J. Hellerstedt, H. Ryu, S. Tang, S.-K. Mo, S. Adam, S. A. Yang, M. S. Fuhrer, and M. T. Edmonds, *Electric-Field-Tuned Topological Phase Transition in Ultrathin $Na_3Bi$*, Nature **564**, 7736 (2018).

[33] A. A. Soluyanov and D. Vanderbilt, *Computing Topological Invariants without Inversion Symmetry*, Phys. Rev. B (2011).

[34] Q. Wu, S. Zhang, H.-F. Song, M. Troyer, and A. A. Soluyanov, *WannierTools: An Open-Source Software Package for Novel Topological Materials*, Comput. Phys. Commun. **224**, 405 (2018).

[35] M. Taherinejad, K. F. Garrity, and D. Vanderbilt, *Wannier Center Sheets in Topological Insulators*, Phys. Rev. B **89**, 11 (2014).

[36] Q. Gu, S. K. Pandey, and R. Tiwari, *A Computational Method to Estimate Spin–Orbital Interaction Strength in Solid State Systems*, Comput. Mater. Sci. **221**, 112090 (2023).

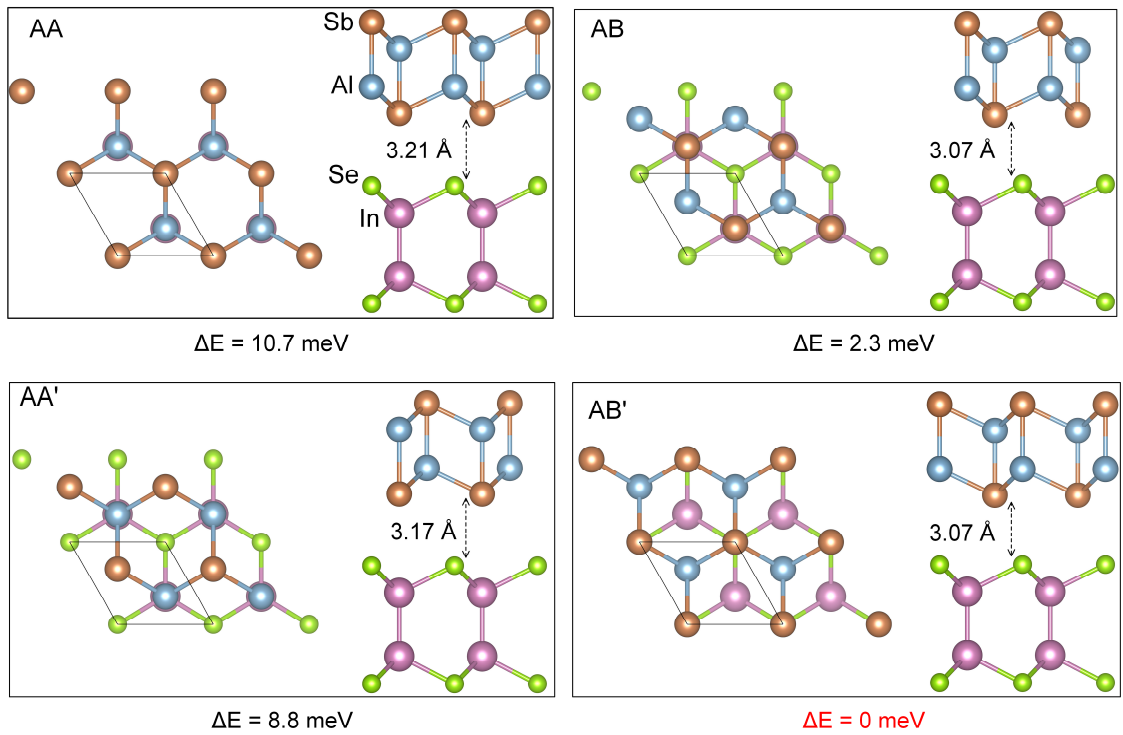

Fig. 1 Top view and side view of the relaxed geometries of AlSb/InSe heterobilayers with the stacking configurations AA, AB, AB', and AA'. The primitive unit cell is denoted by a rhombus. The interlayer distance and the energy difference per cell relative to the AB' configuration are shown.

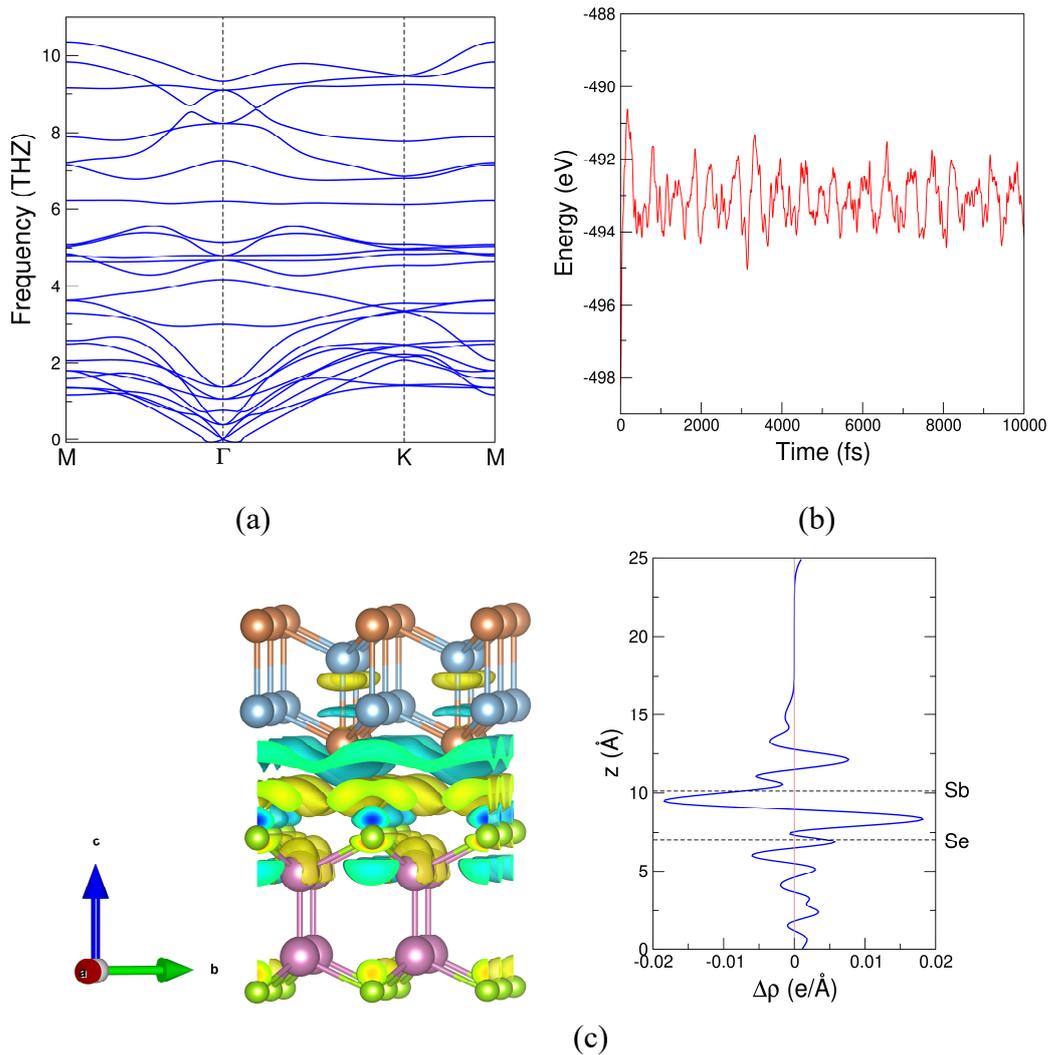

Fig. 2 Phonon dispersion (a), change of energy with time obtained from the MD simulation at 300 K in a canonical ensemble (b), and charge density difference (c) for the AB'-stacked AlSb/InSe heterobilayer. Cyan and yellow represent the depletion and accumulation regions of electrons, respectively.

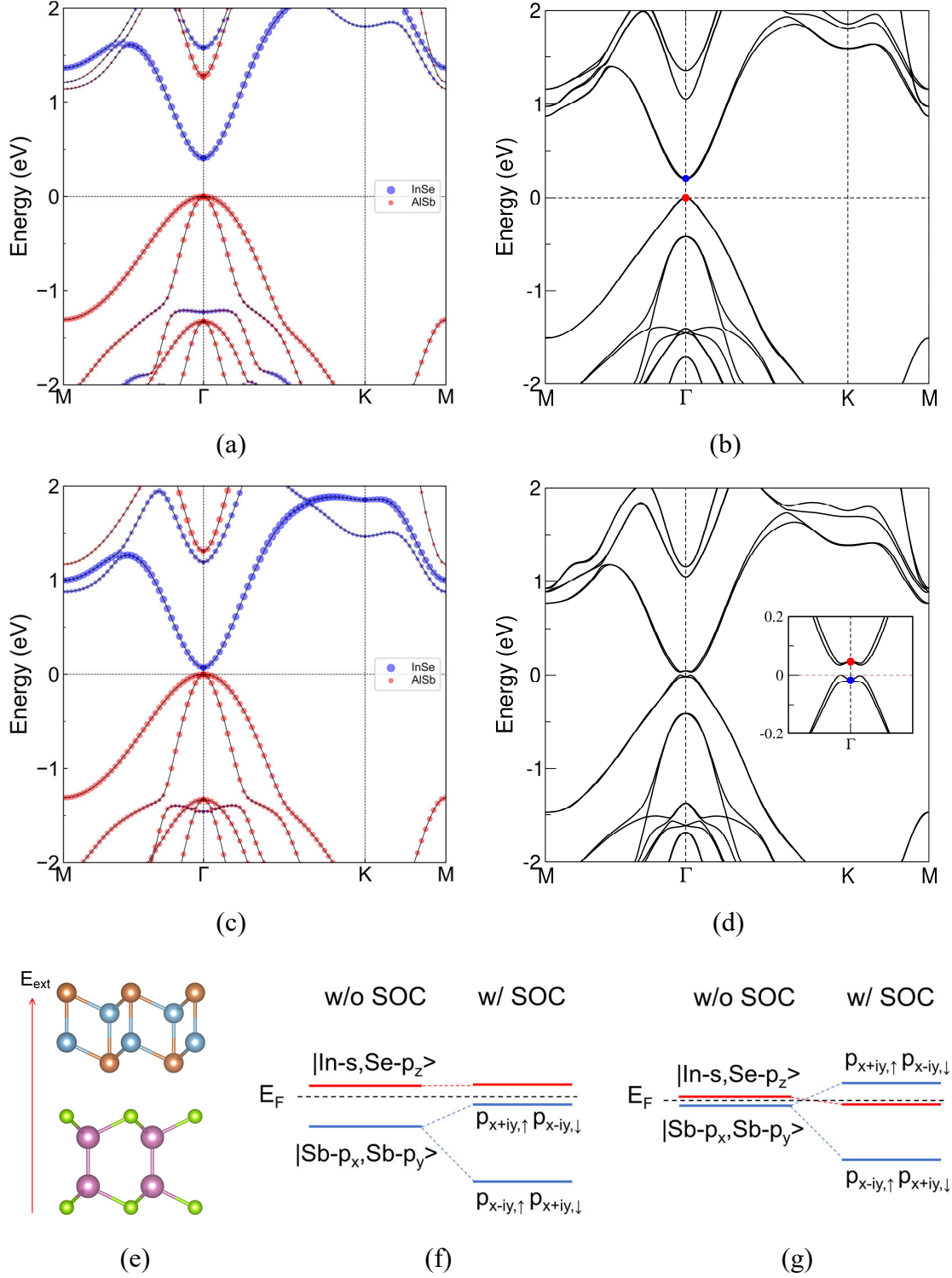

Fig. 3 Electronic band structure of the AB'-stacked AlSb/InSe heterobilayer calculated using hybrid functional in the absence of external electric field without (a) and with (b) SOC and under 0.4 V/Å external electric field without (c) and with (d) SOC (inset: enlarged view around the Fermi level). The top of valence band is set to zero. Red (blue) marks the states contributed by AlSb (InSe). (e) The direction of applied external

electric field is shown. Schematic diagram of the orbitals' evolution at the Γ point: (f) no electric field and (g) $E_{ext}$ = 0.4 V/Å.

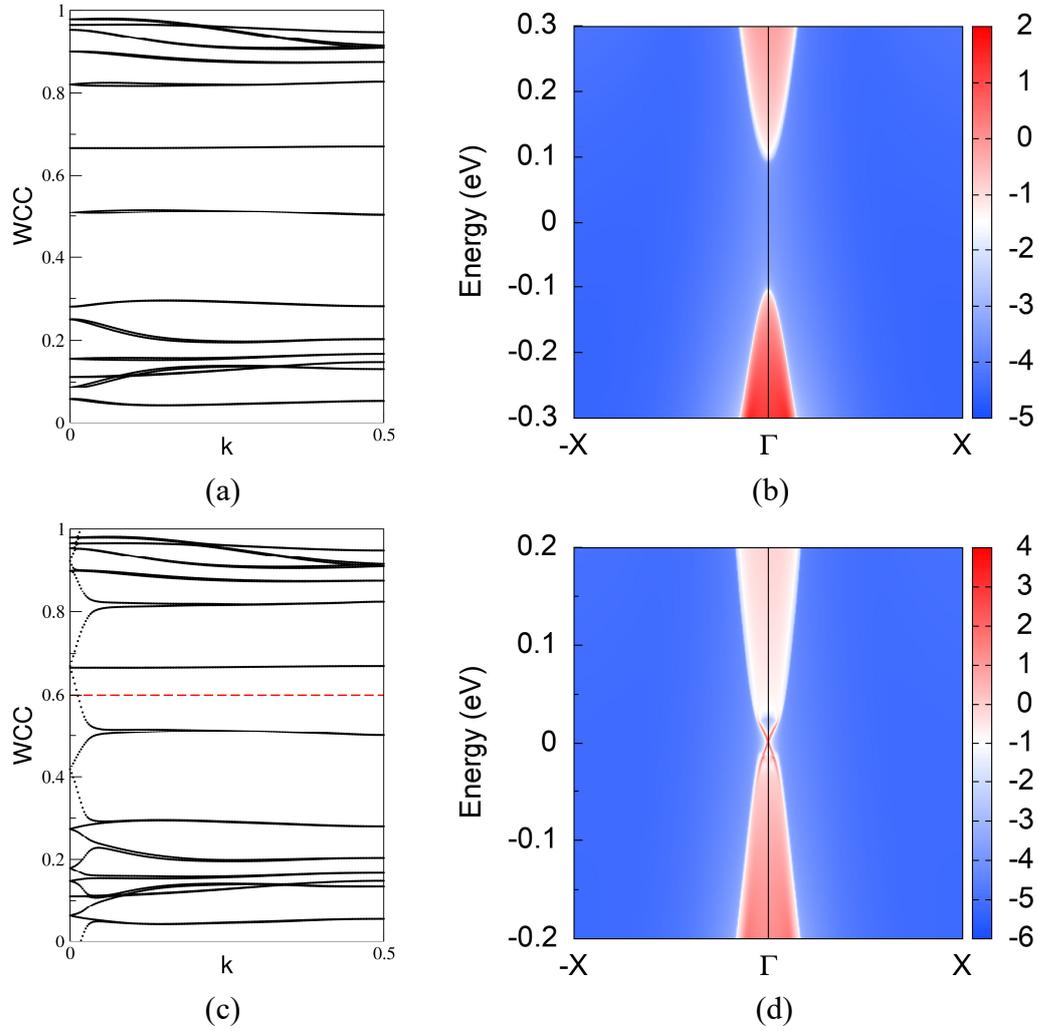

Fig. 4 Evolution of the WCC and edge spectrum (zigzag edge) of the AB'-stacked AlSb/InSe heterobilayer in the absence of external electric field [(a) and (b)] and under $E_{ext}$ = 0.4 V/Å [(c) and (d)].

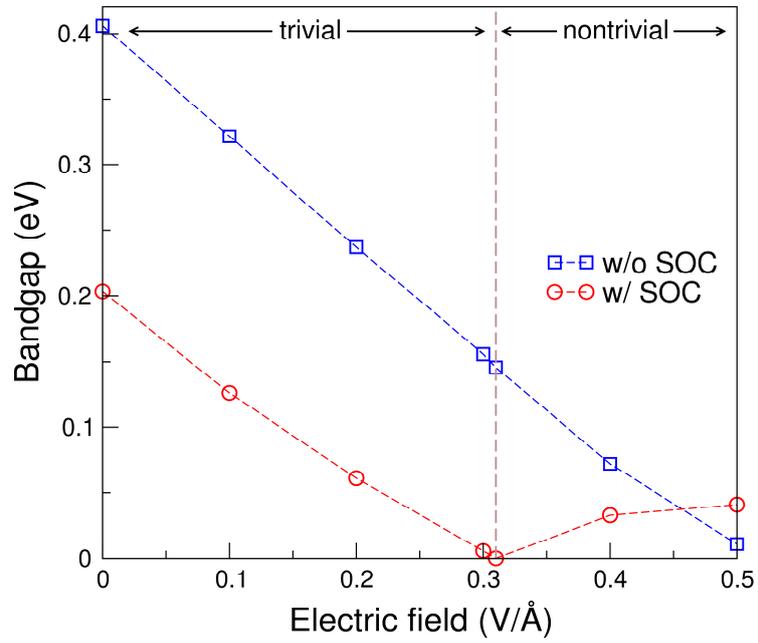

Fig. 5 Bulk bandgap of the AB'-stacked AlSb/InSe heterobilayer as a function of the magnitude of external electric field. A topologically trivial to nontrivial phase transition occurs at 0.31 V/Å.

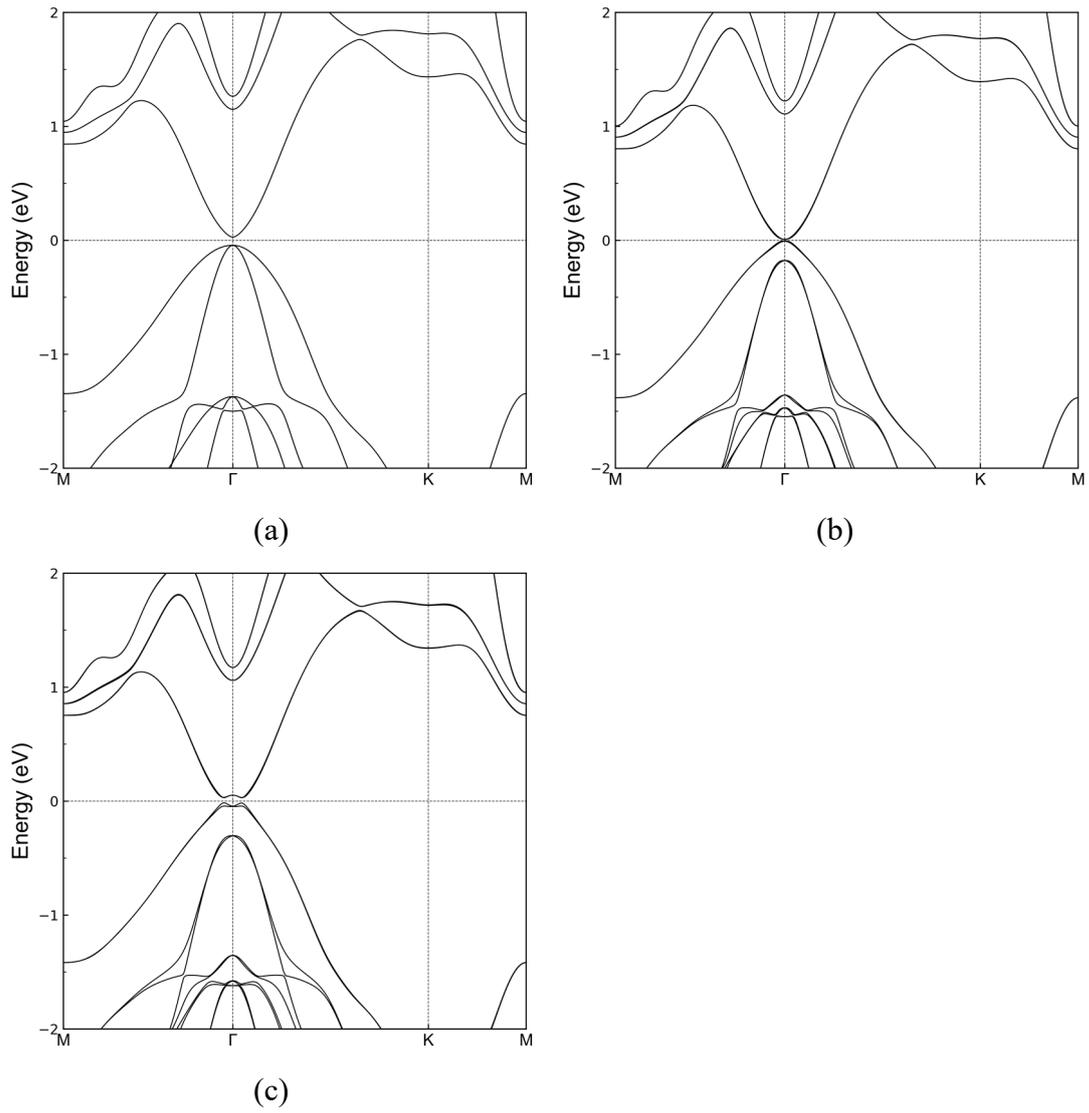

Fig. 6 Electronic band structure of the AB'-stacked AlSb/InSe heterobilayer under $E_{ext}$ = 0.4 V/Å from the tight-binding model: (a) no SOC, (b) $\lambda_{Sb}$ = 0.2 eV, (c) $\lambda_{Sb}$ = 0.4 eV.

**Supporting Material:**

**Electric-field-modulated topological phase transition in AlSb/InSe heterobilayers**

D. Q. Fang[1*], D. W. Wang[2,3]

[1]MOE Key Laboratory for Nonequilibrium Synthesis and Modulation of Condensed Matter, School of Physics, Xi'an Jiaotong University, Xi'an 710049, China

[2]School of Microelectronics & State Key Laboratory for Mechanical Behavior of Materials, Xi'an Jiaotong University, Xi'an 710049, China

[3]Key Lab of Micro-Nano Electronics and System Integration of Xi'an City, Xi'an Jiaotong University, Xi'an 710049, China

*fangdqphy@xjtu.edu.cn

Figures S1-S5.

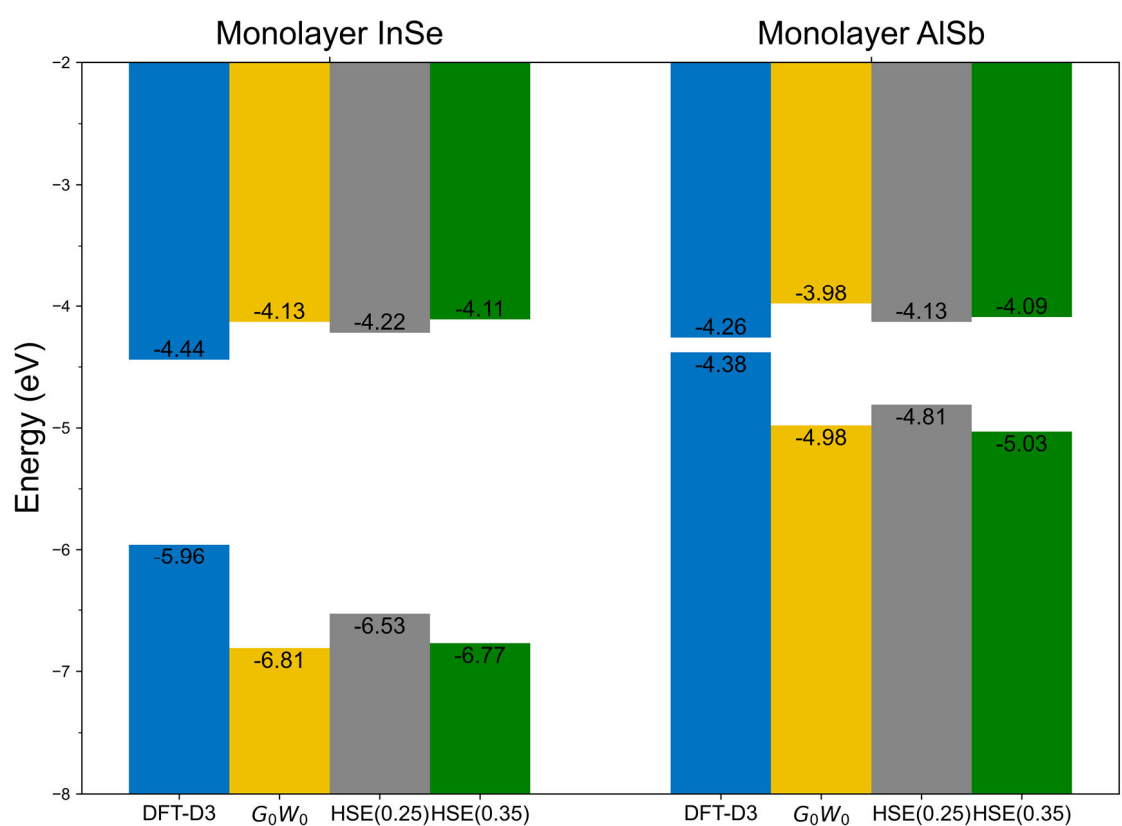

Figure S1. Comparison of the positions of band edges of InSe and AlSb monolayers relative to vacuum calculated using DFT-D3, $G_0W_0$, and HSE06 hybrid functionals with 25% and 35% Hartree-Fock (HF) exchanges, respectively.

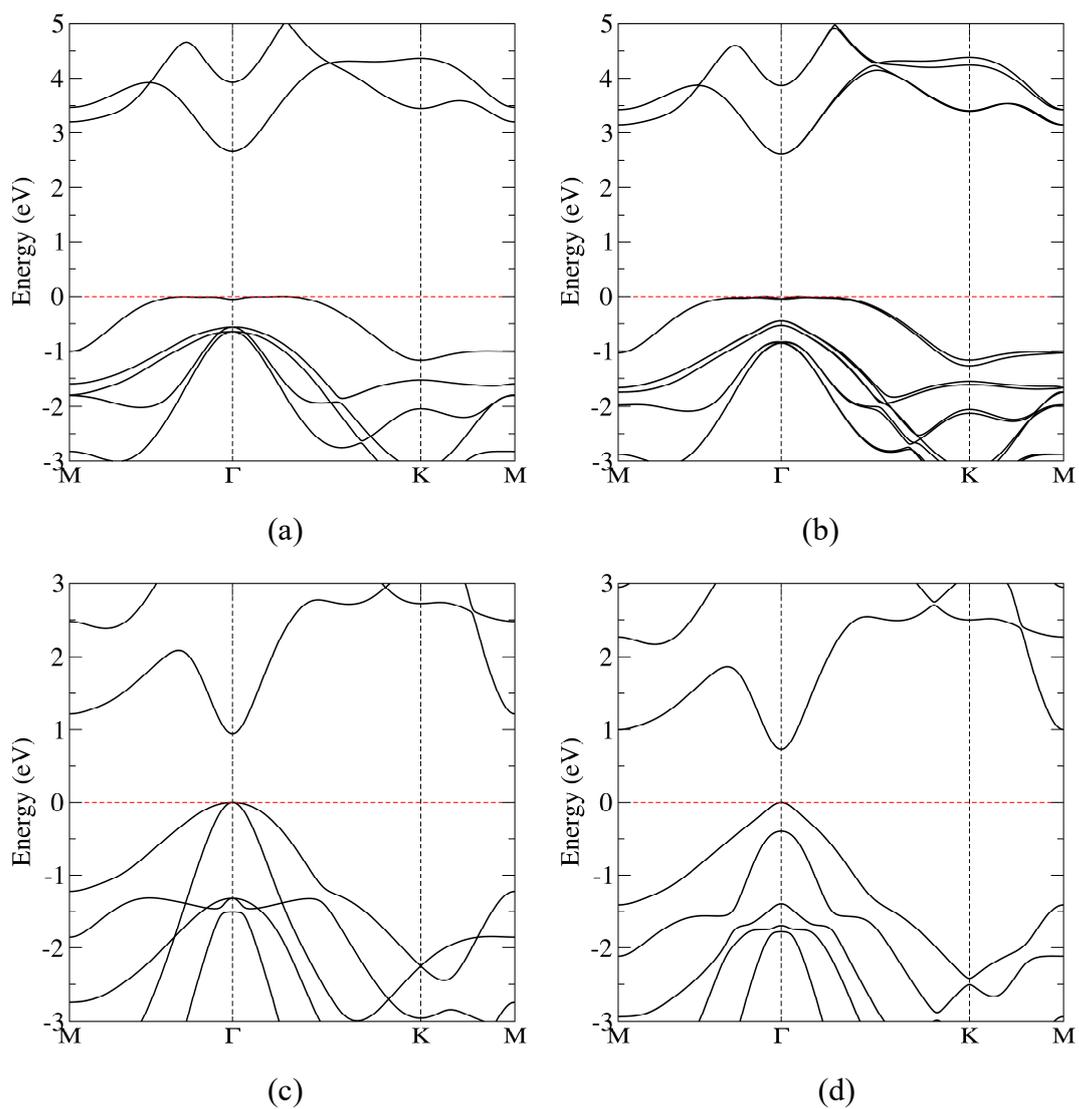

Figure S2. Electronic band structure of monolayer InSe without (a) and with (b) SOC and monolayer AlSb without (c) and with (d) SOC calculated using the HSE06 hybrid functional with 35% HF exchange. The top of valence band is set to zero.

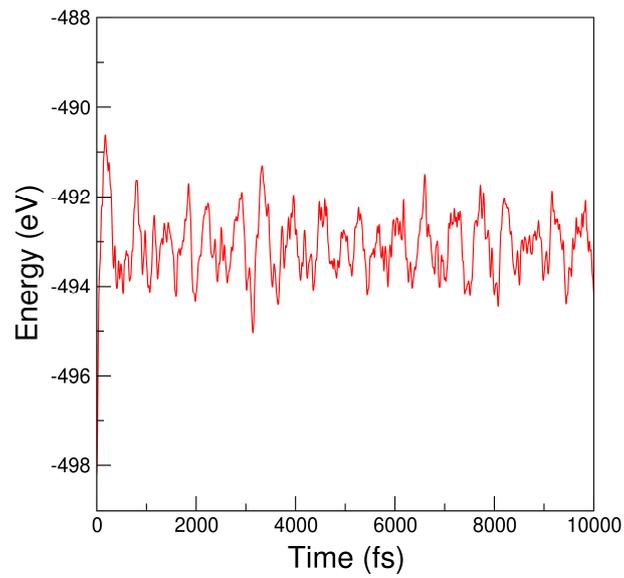

(a)

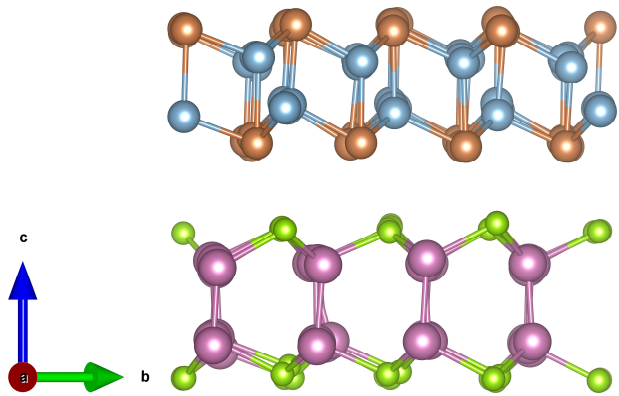

(b)

Figure S3. Change of energy with time obtained from the MD simulation at 300 K in a canonical ensemble (a) and structure snapshot at 10 ps (b) for the AB'-stacked AlSb/InSe heterobilayer.

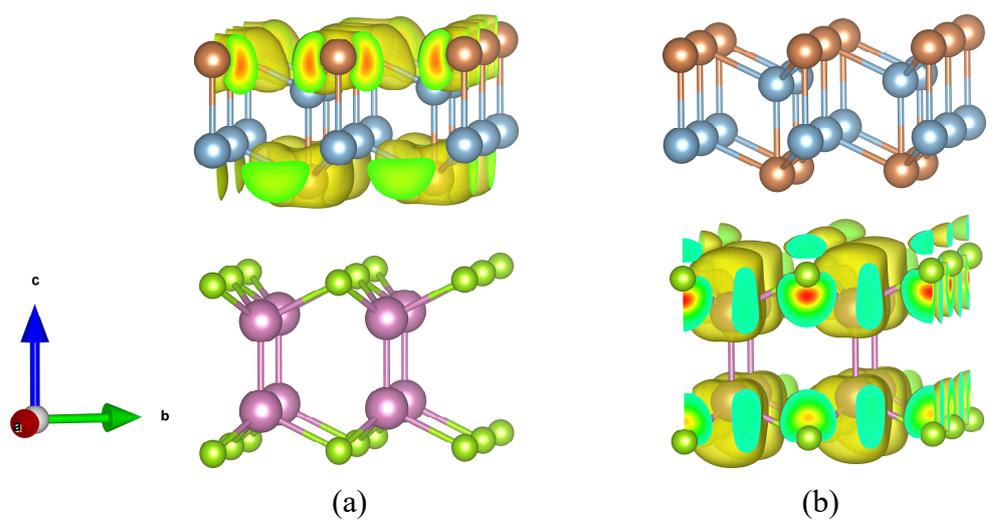

Figure S4. VBM (a) and CBM (b) states of the AB'-stacked AlSb/InSe heterobilayer.

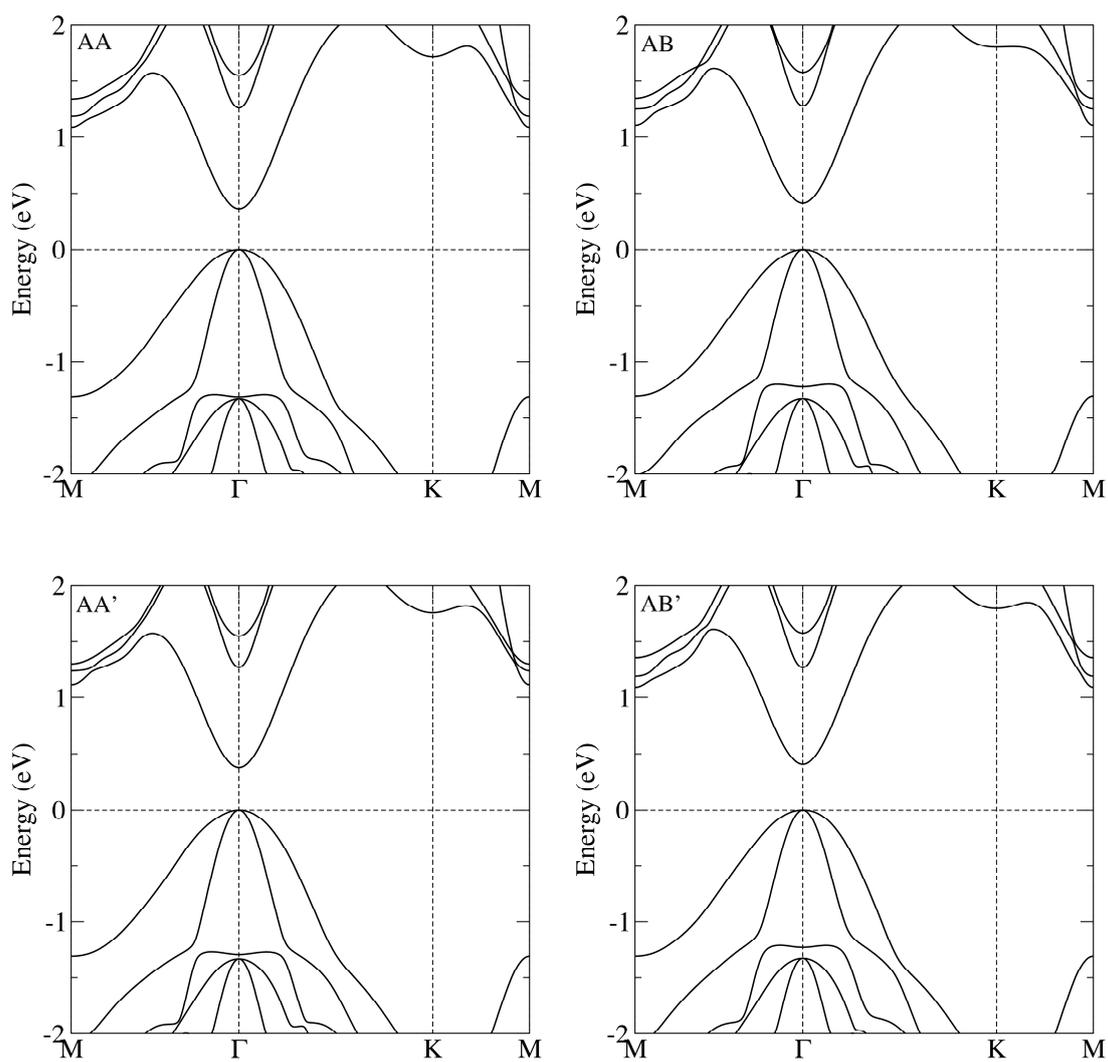

Figure S5. Hybrid functional band structure without SOC for the AA, AB, AB', and AA' stacking configurations calculated via Wannier interpolation. The top of valence band is set to zero.